\documentclass{aa}
\usepackage{graphicx}
\usepackage{natbib}
\usepackage{amsmath}
\bibpunct{(}{)}{;}{a}{}{,}

\newcommand{\kms}{km~s$^{-1}$}

\newcommand{\rmax}{$R_{max}$}
\newcommand{\rmin}{$R_{min}$}

\begin{document}
\headnote{Letter to the Editor}

\title{Evidence for an inner molecular disk around massive Young
  Stellar Objects \thanks{Based on observations collected at the
    European Southern Observatory at La Silla and Paranal, Chile (ESO
    Programmes 68.C-0652 and 69.C-0448)}}

   \author{A.~Bik\inst{1} \and W.~F.~Thi\inst{1}}

   \offprints{A. Bik (bik@science.uva.nl)}

   \institute{Astronomical Institute ``Anton Pannekoek'',
           University of Amsterdam, Kruislaan 403, 1098 SJ Amsterdam,
           The Netherlands
             }

   \date{Received; accepted}

\authorrunning{A.\ Bik et al.}
\titlerunning{A molecular inner disk around massive YSOs}

\abstract{ We present observations of CO overtone bandhead emission toward
  four massive Young Stellar Objects (spectral type O6--B5). The high
  signal-to-noise ratio $K$-band spectra were obtained with {\em
    VLT-ISAAC} at a resolution $\Delta v$ = 30 \kms, sufficient to
  resolve the bandheads, but not the individual $J$-lines.  We are
  able to explain the shape of the lines by assuming a simple
  isothermal keplerian disk model seen at different inclinations. The
  gas temperature ranges from 1500 to 4500~K and the CO column density
  is between 0.1 and 4 $\times$ 10$^{21}$ cm$^{-2}$.  The emission
  probably arises within the first few astronomical units of the disk,
  consistent with the high gas temperature.  Our results indicate that
  molecules can survive close to a hot star and suggest that dense
  ($n_{\rm H}>10^{10}$ cm$^{-3}$) inner disks may be relatively common
  at an advanced stage of high-mass star formation.  \keywords{Stars:
    early-type, formation, circumstellar matter, molecular processes,
    Line: profiles }}\maketitle

%_______________________________________________________________
   
\section{Introduction}

Disks are found around many young low-mass stars as a natural
byproduct of angular momentum conservation during the star formation
process. Observational evidence for disks around their more massive
counterparts remains sparse, although they may be essential during the
formation of high-mass stars \citep{Churchwell02}.  Extended disks
have been found around massive young stellar objects (YSOs) in the Hot
Core phase
\citep[e.g.][]{Shepherd01Science,Beltran04,Garay99,Chini04}, albeit
those millimeter observations solely probe gas beyond 1000~AU. In the
later Herbig Be stage, disks have only been recently discovered
\citep{Fuente03}.

Evidence for the presence of a gaseous disk around the early B star
\object{NGC2024-IRS2} is presented by \citet{Lenorzer04}.  They argue
that the infrared excess observed toward \object{NGC2024-IRS2} is best
explained by the emission from a dense gaseous disk as close as 0.3~AU
to the star.

At the high densities and temperatures (1000--2000~K) characteristic
of disks at 0.1--5~AU around young stars, molecules are expected to be
sufficiently excited to produce a rich ro-vibrational spectrum in the
near- (overtone transitions $\Delta v = \pm2$) and mid-infrared
(fundamental transitions $\Delta v = \pm 1$). CO first overtone
emission is most likely emitted by a disk and/or a wind
and has been detected and analyzed by, e.g.,
\citet{Scoville79,Geballe87,Carr89,Chandler93,Chandler95} toward a
number of other high- and low-mass stars.  Their studies were, however,
hampered by the small number of observed $J$-lines or a low
resolving power and moderate signal-to-noise.

\citet{Brgspec04} examine $K$-band spectra of  a sample of massive YSOs
associated with ultra-compact H{\sc ii} regions (UCHII). Here, we present
a simple disk model that describes the CO bandhead emission detected
in four of these objects. The observations are presented in
Sect.~\ref{sec:obs}. The disk model and the results are shown in
Sect.~\ref{sec:model}. In Sect.~\ref{sec:discussion} we discuss the
location and origin of the CO. The implications of our findings on the
process of high-mass star formation are summarized in Sect.~\ref{sec:conclusion}.
% ---------------------------------------------------------------------

\begin{figure*}
\centering \resizebox{\hsize}{!}{\includegraphics{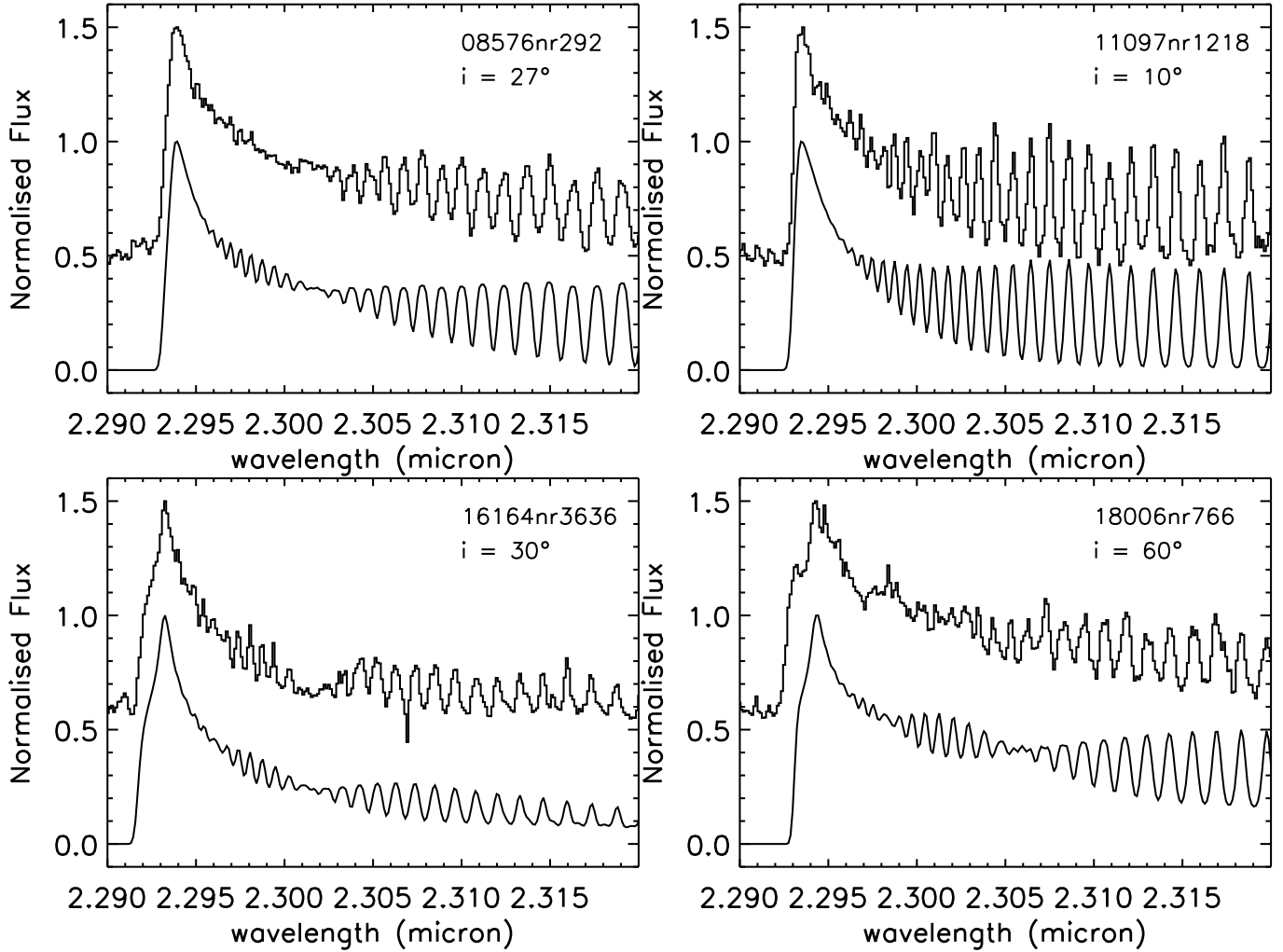}}
   \caption{$K$-band observed (upper spectra) and modeled (lower
     spectra) spectra of the CO 2--0 bandhead. The name of the object
     and the derived inclination is displayed.  By definition $0
     \degr$ inclination corresponds to a disk seen pole-on. }
   \label{fig:COfits}
\end{figure*}
% ---------------------------------------------------------------------

\section{Observations}\label{sec:obs}

High signal-to-noise ratio long-slit $K$--band spectra of four massive
YSOs have been obtained in 2002 with the {\em ISAAC} spectrometer
mounted on the {\em VLT}. A narrow slit (0.3\arcsec) is used,
resulting in a resolving power of $R$ = 10\,000 ($\Delta v$ = 30 km
s$^{-1}$). The spectra were reduced using standard procedures. The
wavelength calibration is based on the telluric OH emission
line spectra. Telluric standard stars of spectral type A0V, observed under identical
sky conditions as the science targets, were used
to correct for the telluric absorption lines. A detailed description and
analysis of the $K$-band spectra can be found in \citet{Brgspec04}.
The properties of the objects are summarized in  Table \ref{tab:fitparams}.
%A
%summary of the properties of the objects is given in table
%\ref{tab:fitparams}.

% ---------------------------------------------------------------------
\begin{figure*}
\centering \resizebox{0.95\hsize}{!}{\includegraphics{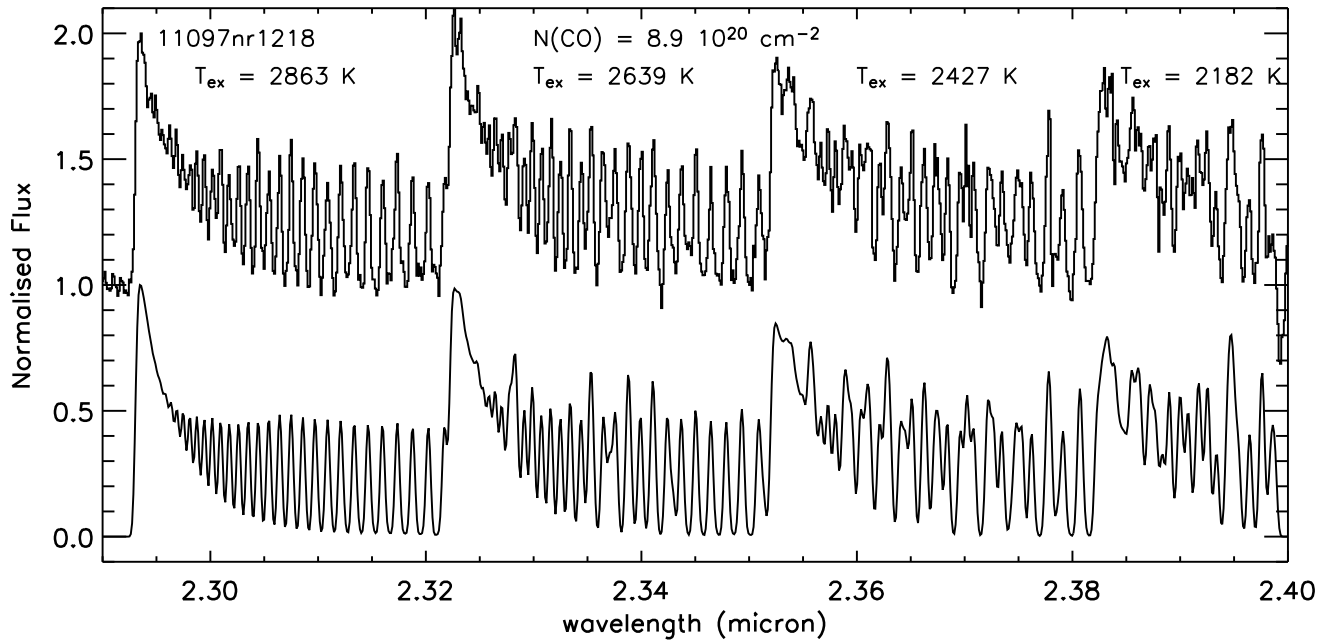}}
 \caption{Fit of the complete observed CO bandhead spectrum of
 \object{11097nr1218}. The {\em upper} curve is the observed spectrum
 whereas the {\em lower} one is the model spectrum.}
   \label{fig:COfit4bands}
\end{figure*}

% ---------------------------------------------------------------------

\begin{table*}
\begin{center}

\begin{tabular}{lllllllllll} 
\hline
\noalign{\smallskip}

\multicolumn{1}{c}{IRAS}&\multicolumn{1}{c}{Object}&\multicolumn{1}{c}{Mass}&\multicolumn{1}{c}{$d$}&\multicolumn{1}{c}{$T_{\rm ex}$}& \multicolumn{1}{c}{$N$(CO)} &\multicolumn{1}{c}{$i$}& \multicolumn{1}{c}{$R_{\rm min}$} & \multicolumn{1}{c}{$R_{\rm max}$} & \multicolumn{1}{c}{$M$(CO)} & \multicolumn{1}{c}{$\chi^2_{\nu}$} \\
  &  &\multicolumn{1}{c}{(M$_{\sun}$)}  &\multicolumn{1}{c}{(kpc)}&\multicolumn{1}{c}{(K)} &\multicolumn{1}{c}{(cm$^{-2}$)} &\multicolumn{1}{c}{($\degr$)}& \multicolumn{1}{c}{(AU)} &\multicolumn{1}{c}{(AU)} & \multicolumn{1}{c}{(M$_{\sun}$)} & \\ 
\hline
\noalign{\smallskip}
\object{16164-5046}    & \object{16164nr3636}    & 30 & 3.6    & 4480    & $3.9 \times 10^{20}$   & 30$^{+ 6}_{-23}$ & 3.1$^{+ 18.2}_{-2.9}$ & 3.2$^{+123.5}_{-3.0}$  & 2.0 $\times 10^{-9}$  & 0.99\\
\object{11097-6102}    & \object{11097nr1218}    & 11 & 2.8    & 2710    & $8.9 \times 10^{20}$   & 10$^{+ 10}_{-5}$  & 0.6$^{+0.1}_{-0.4}$  & 3.3$^{+ 13.5}_{-3.1}$   & 2.6 $\times 10^{-7}$    & 1.51\\
\object{18006-2422}    & \object{18006nr766}     & 11  & 1.9    & 1800    & $1.0 \times 10^{21}$   & 60$^{+ 20}_{-30}$ & 0.2$^{+ 0.24}_{-0.06}$ & 1.9$^{+2.5}_{-1.6}$   & 6.2 $\times 10^{-8}$    & 1.72\\
\object{08576-4334}    & \object{08576nr292}     & 6  & 0.7    & 1660    & $3.9 \times 10^{21}$   & 27$^{+ 2}_{-14}$  & 0.2$^{+1.1}_{-0.1}$ & 3.6$^{+70}_{-3.4}$       & 8.4 $\times 10^{-7}$    & 2.01\\

\noalign{\smallskip}
\hline
\end{tabular}  

\caption{Stellar properties and fit parameters of the first ($v=2-0$)
  bandhead. The masses of the objects (column 3) are based on spectral
  type estimates, determined from their position in the ($K,J-K$)
  color-magnitude diagram \citep{Brgspec04}. The reference of the
  distances (column 4) are given in \citet{Brgspec04}.  The errors on
  the inclination reflect the range of inclinations where a good fit
  was possible.  The errors on \rmin\ and \rmax\ are the corresponding
  values for these inclinations.The errors on $T_{ex}$ and $N(CO)$ are
  dominated by systematic instead of statistical
  errors. \label{tab:fitparams}}
 \end{center}

\end{table*}

%The errors of
%  $T_{ex}$ and $N(CO)$ are dominated by systematic instead of
%  statistical errors.

% ---------------------------------------------------------------------

%are taken from \citet{Brgspec04} and based on their position in the ($K,J-K$) colour-magnitude diagram.

%O6--O8V (
%Early B (
%mid-B   (
%mid-B   (

\section{Disk model}\label{sec:model}

Continuum subtracted and normalized first-overtone bandhead spectra
are shown in Fig. \ref{fig:COfits} (upper spectra). Four overtones are
detected but only the first one is displayed. The shape of the
bandhead varies from source to source.  The rise of the first bandhead
is sharp in the spectrum of \object{08576nr292} and
\object{11097nr1218} whereas in that of \object{16164nr3636} and
\object{18006nr766} a blue wing is clearly present.

The expected profile from a spherical stellar wind has a flat-topped
shape and cannot account for the wing seen in two of the objects. 
The favored model to explain the blue wing consists of a keplerian
disk and/or a disk-wind \citep[e.g.][]{Chandler95}, although the disk
model provided the best match.  In this paper, we explore the possibility that
the CO bandhead emission arises solely from a small disk in keplerian
rotation around the massive star. We generate synthetic CO bandhead
spectra using a standard parametric disk model for a range of gas
temperatures, column densities, turbulent velocities and disk viewing
angles \citep[cf.][ for a detailed description]{Kraus00}. In contrast
to standard disk models, the disk is assumed to be entirely
isothermal.  The molecular data (line frequencies and Einstein
coefficients) are taken from the database of \citet{Chandra96}. At
temperatures above 1500~K, needed to populate the higher vibrational
levels, no dust grains can survive and, therefore, no dust opacity is
included.  The population of the CO rotational levels within each
vibrational level is assumed to be in local thermodynamic equilibrium
(LTE), at the same temperature $T_{\rm ex}$ as the local vibrational
temperature. Each bandhead is characterized by a specific vibrational
temperature. The downhill simplex method is used to determine the
model parameters: the CO total column density $N$(CO), the excitation
temperature $T_{\rm ex}$, the disk inner and outer radius $R_{\rm
min}$ and $R_{\rm max}$ (Table \ref{tab:fitparams}) corresponding to
the best fitting model for each object (see the bottom spectra in each
panel of Figs.  \ref{fig:COfits} and \ref{fig:COfit4bands}). The
stellar mass and the inclination are degenerate parameters.  The
parameters are best constrained by the first bandhead which is not
contaminated by the $P$-branch of the other bandheads nor by Hydrogen
Pfund lines.  For \object{11097nr1218}, we were able to fit the four
bandheads (Fig.~\ref{fig:COfit4bands}) which shows that $T_{\rm ex}$
is lower for higher $v$ (see next section).

\section{Results and Discussion}\label{sec:discussion}

We have observed the CO bandheads in a sample of 15 massive YSOs and
five show these lines in emission \citep{Brgspec04}. In this
\emph{Letter}, four of these objects are modeled and are well
described by optically thin CO gas located in a keplerian disk. From
the quality of the fits, it is clear that additional emission from a
disk-wind is not required.

Our results show that the CO gas is located within a few AU of the
central star (Table \ref{tab:fitparams}). The maximum distance where
the CO bandheads are emitted depends on the inclination, which is not
a well constrained parameter, but even for a wide range of possible
inclinations the hot CO molecules are located within a few AU from the
star.

In the absence of extinction by dust grains, the CO molecules should
be photodissociated by the stellar ultraviolet photons.  However, the
derived column densities (10$^{20}$ -- 10$^{21}$ cm$^{-2}$) are well
above the required value for the CO molecules to self-shield
\citep[$N$(CO) $\sim 10^{15}$cm$^{-2}$,][]{vanDishoeck88}. Moreover,
gas phase chemical models show that CO molecules can rapidly form in
the gas phase to compensate for their destruction.

The CO column densities in the four objects only differ by at most a
factor of 10. This reflects the narrow parameter range in density and
temperature at which the CO first-overtone bandheads are emitted. A
steep density and/or temperature gradient can result in a very narrow
region in the disk where the CO first-overtone is emitted. This is
reflected in the value of \rmin\ and \rmax\ in e.g.
\object{16164nr3636}, where only a narrow ring in the possibly much
more extended disk is producing the CO bandheads.

The derived excitation temperatures range from 1700 to 4500~K. This
temperature may not be equal to the gas kinetic temperature for the
high-$v$ transitions since the critical density is increasing with $v$
($n_{\rm H} >$~10$^{10} \mbox{cm} ^{-3}$). In addition, the fit to all four
bandheads shows that $T_{\rm ex}$ is lower for higher $v$ transitions.
Therefore, the population of the high $v$ levels is likely
sub-thermal.  The excitation temperature shows a positive trend with
increasing effective temperature of the central star (Table
\ref{tab:fitparams}). One possibility is that the diluted
near-infrared radiation field pumps the higher $v$-levels. Likewise,
UV fluorescence will advantageously populate the high $v$-levels.
Alternatively, the earlier type stars have higher luminosity which
results in an increased heating by photo-electrons. The later
explanation is favored because both radiation pumping mechanisms (UV
and near-IR) will preferably populate the higher rather than the lower
$v$-levels, which is not corroborated by the data. However, only a
detailed model that includes all the CO levels and possible pumping
mechanisms will give a definitive answer.

Two scenarios can explain the presence of an inner disk at the stage
of massive star formation where the circumstellar environment is being
dispersed by the stellar wind and radiation.  

According to the first scenario, the disk material may originate from
the rapidly rotating star itself. However, only the wind of a B[e]
supergiant can produce a disk with sufficiently high densities to emit
CO bandheads \citep{McGregor88,Kraus03}. The stellar characteristics
exclude the possibility that they are supergiants \citep{Brgspec04}.

Alternatively, the inner disk may be the remnant of a massive and
extended (up to 10\,000~AU) accretion disk detected in the early phase
of star formation
\citep[e.g.][]{Shepherd01Science,Beltran04,Garay99,Chini04}. In the
model of disk dispersal by \citet{Hollenbach94}, the photoevaporation
of the outer part is faster than the dispersion of the inner region by
stellar winds. This model naturally explains both the presence of
inner disks and the absence of large outer disks in most Herbig Be
stars \citep[e.g.][]{Natta00,Fuente03}.

\section{Conclusions \& Prospects}\label{sec:conclusion}

We have modeled the CO first-overtone bandhead emissions of 4 massive
YSOs. The different profiles of the CO bandheads are all well
explained by a simple keplerian disk. Regardless of the gas
excitation mechanism, the information on the gas kinematics given by
the shape of the CO bandheads shows that the CO is located close to
the star (0.1--5~AU).

The CO overtone emission likely arises in a narrow torus where the gas
temperature is between 1500 and 4500~K. This range of temperature
reflects the difference in  effective temperature of the central objects.
The decrease in excitation temperature for one object from the
$v=$~2--0 bandhead to the $v=$~5--3 bandhead indicates that the higher
$v$-levels are sub-thermally populated. Excitation mechanisms other
than LTE, like UV or infrared pumping and photo-electric heating
cannot be excluded by the simple model used here.

The detected disks may be the remnants of massive accretion disks. In
particular, it is the first evidence of a disk around a
young O6--O8 star, suggesting that these objects are recently
formed as the disks are not yet dispersed by the stellar FUV photons
and stellar wind. 

%young massive stars.

Spatially resolved observation of CO bandhead emissions with {\em
  Amber} at the {\em VLT} interferometer will provide further
  constraints on the size and the geometry of the CO emitting
  region.

\begin{acknowledgements}  
  WFT is supported by NWO grant 614.041.005. The authors thank Rens
 Waters, Lex Kaper and Doug Johnstone for fruitful discussion and the VLT
 staff for support and help with the observations.
\end{acknowledgements}

%\bibliographystyle{aa}
% change path...
%\bibliography{/home/abik/Papers/arjan}
%\bibliography{/scratch/thi/bik/arjan}
%\bibliography{}

\end{document}